\documentclass[aps,prl,groupeaddress,color,supperscriptaddress,showpacs,footinbib,twocolumn,showkeys,final]{revtex4}
\usepackage{graphics,graphicx,color}
\def\tauphi{$\tau_{\phi}$ }

\begin{document}

\title{Effect of Disorder on the Quantum Coherence in Mesoscopic Wires}
\author{Y.~Niimi$^{1,\dagger}$, Y.~Baines$^{1}$, T.~Capron$^{1}$, D.~Mailly$^{2}$, F.-Y.~Lo$^{3,\ddagger}$, A.~D.~Wieck$^{3}$, T.~Meunier$^{1,*}$, L.~Saminadayar$^{1,4,5,*}$ and C.~B\"auerle$^{1,*}$}
\affiliation{$^{1}$Institut  N\'{e}el, CNRS, B.P. 166, 38042
Grenoble Cedex 09, France} \affiliation{$^{2}$Laboratoire de
Photonique et Nanostructures, route de Nozay, 91460 Marcoussis,
France} \affiliation{$^{3}$Lehrstuhl f\"{u}r Angewandte
Festk\"{o}rperphysik, Ruhr-Universit\"{a}t, Universit\"atsstra{\ss}e
150, 44780 Bochum, Germany} \affiliation{$^{4}$Universit\'{e} Joseph
Fourier, B.P. 53, 38041 Grenoble Cedex 09, France}
\affiliation{$^{5}$Institut Universitaire de France, 103 boulevard
Saint-Michel, 75005 Paris, France}

\date{1 June 2009}

\begin{abstract}
We present phase coherence time measurements in quasi-one-dimensional 
mesoscopic wires made from high mobility two-dimensional electron gas. 
By implanting gallium ions into a
GaAs/AlGaAs heterojunction we are able to vary the diffusion
coefficient over 2 orders of magnitude. We show that in the
diffusive limit, the decoherence time follows a power law as a
function of diffusion coefficient as expected by theory. When the
disorder is low enough so that the samples are semi-ballistic, we
observe a new and unexpected regime in which the phase
coherence time is independent of disorder. In addition, for all
samples the temperature dependence of the phase coherence time
follows a power law down to the lowest temperatures without any sign
of saturation and this strongly suggests that the frequently observed low
temperature saturation is \textit{not} intrinsic.
\end{abstract}

\maketitle

Decoherence is the process by which a quantum system looses phase
coherence by getting entangled with its environment. For a metallic
system, the dominant mechanisms for decoherence of electrons are
inelastic scattering events due to collisions with other electrons,
phonons, and also extrinsic sources such as magnetic impurities or
two level systems. Whereas at temperatures above a few Kelvin the
electron-phonon interactions are dominant, at low temperatures and
without extrinsic sources of decoherence, the leading decoherence
process is due to electron-electron interactions and has been at the
center of a controversy over the last decade: the \textquotedblleft
standard\textquotedblright~theory of electron coherence in metals by
Altshuler, Aronov and Khmelnitski~(AAK)~\cite{AAK_82} predicts a
diverging dephasing time with vanishing temperature, while an
alternative theory proposed by Golubev and Zaikin (GZ) predicts a
saturation of the dephasing time at low temperatures
\cite{GZ_prl_98}. Saturation of the dephasing time down to absolute
zero raises the question of the breakdown of the Fermi liquid theory
and, as such, is of basic importance for our understanding of
metallic conductors.

The experiments by Mohanty et al.~\cite{mohanty_prl_97}, showing a
systematic saturation of the phase coherence time $\tau_{\phi}$ at low
temperature have triggered a large amount of studies, 
both from experimental and theoretical points of view.
These studies aimed to show whether 
the saturation of $\tau_{\phi}$ 
at low temperature is either intrinsic~\cite{GZ_prl_98,mohanty_prl_97}, 
or due to any other extrinsic mechanisms~\cite{imry_epl_99,zawa_prl_99,natelson_prl_01,birge+pierre_prl_02,pierre_prb_03,schopfer_prl_03,gershenson_prl_06}.
It should be stressed, however, that for all these
experiments only the \emph{temperature} dependence of $\tau_{\phi}$ has
been investigated. On the other hand, it has been shown that this
approach is by far insufficient, as the presence of a tiny amount of
magnetic impurities with given Kondo temperature can always account
for the experimentally observed saturation of
$\tau_{\phi}$~\cite{bauerle_prl_05,mallet_prl_06,sami_physicaE_07}. 
It is thus utterly important to find
another parameter to be able to discriminate between the different
scenarios.

It is our purpose to break this impasse by taking a new approach.
Instead of studying the temperature dependence of \tauphi we exploit
another property which should shine new light on this problem and
which has been almost unexplored so far: the dependence of the
phase coherence time on the disorder, or, in other words, on the
diffusion coefficient $D$. Some attempts to measure this dependence
have been realized in metallic systems~\cite{ovadyahu_83,Lin_2001}, but in this
case it is very difficult to vary $D$ in a controlled way over a
wide range. Another study has been performed on two-dimensional
electron systems~\cite{Noguchi_JAP_96}, but in that work the phase
coherence time was already saturating at 2 K in their cleanest
samples. Those works thus do not allow to draw any clear conclusions
about the dependence of the phase coherence on disorder.

In this Letter, we report on measurements of the electronic phase
coherence time $\tau_{\phi}$ in wires made from a 
two-dimensional electron
gas (2DEG) where the diffusion coefficient 
$D$ is varied in an
extremely controlled way over 2 orders of magnitude by using an
original ion implantation technique. In the diffusive regime, the
$D$ dependence of the phase coherence time $\tau_{\phi}$ follows a
power law $D^{\alpha}$ with $\alpha$ close to $1/3$. More
surprisingly, when entering the semi-ballistic regime, 
the phase coherence time becomes independent of disorder. 
Moreover, for all
samples the temperature dependence of $\tau_{\phi}$ follows a power
law down to the lowest temperature without any sign of saturation;
this casts serious doubt that the frequently observed low temperature
saturation could be intrinsic.

The difficulty with measurements of the decoherence in any metal is
the purity of the metal source. A careful choice of the starting
material is of utmost importance as any tiny inclusion of impurities
could lead to an apparent saturation at low temperatures
\cite{sami_physicaE_07}. For this reason we have used a high
mobility GaAs/AlGaAs heterostructure as starting material. These
systems are intrinsically clean and grown in ultrahigh vacuum by
molecular beam epitaxy. Before processing, the 2DEG has an electron
density of $n_{e} = 1.76\times 10^{11}$ cm$^{-2}$ and mobility $\mu
= 1.26\times 10^{6}$ cm$^{2}/$V$\cdot$s at 4 K in the dark. All
lithographic steps are performed using electron beam lithography on
polymethyl-methacrylate resist. Each sample (see inset of
Fig.~\ref{fig2}) consists of $4$ sets of $20$ wires of length $L =
150$ $\mu$m in parallel, and of lithographic width $W = 600$, $800$, 
$1000$ and $1500$ nm. In addition, a Hall bar allows us to measure
the electronic parameters of the 2DEG (i.e. $n_{e}$, $\mu$, the
elastic mean free path $l_{e}$, etc.). The diffusion
coefficient is obtained \textit{via} the relation $D = v_{F}l_{e}/2$
with $v_{F}$ the Fermi velocity. In order to change the diffusion
coefficient of our samples we use a very original technique: first
we write the pattern of each sample on the \emph{same} wafer. We
then place a Focused Ion Beam (FIB) microscope on one sample and
implant locally Ga$^{+}$ ions with an energy of 100~keV. For this
energy, the Ga$^{+}$ ions penetrate only about 50~nm into the GaAs
heterostructure, whereas the 2DEG lies 110 nm below the surface. The
implanted ions create crystal defects in the AlGaAs layer and
modify the intrinsic disorder potential felt by the electrons; note,
however, that in the case of low dopings like here, the
band structure is not changed~\cite{wieck_ss_90}. This affects only
the momentum scattering time and thus the mobility of the itinerant
electrons in the 2DEG \cite{wieck_pss_08}. By varying the
implantation dose for different samples ($10^{8}$ cm$^{-2}$ to
$2.5\times 10^{9}$ cm$^{-2}$) we are able to vary the diffusion
coefficient from 3500 cm$^{2}/$s (unimplanted sample) to 130
cm$^{2}/$s. Magnetoconductance measurements are performed using a
standard ac lock-in technique and a home-made very low noise
preamplifier (0.5 nV$/\sqrt{\rm Hz}$) at room temperature. The
temperature dependence of the resistivity is measured on the Hall
bar and used to check the \emph{actual} temperature of the electrons
of the sample: we find that it follows nicely a $\ln(T)$ law down to
a temperature of 40 mK as expected~\cite{AAK_82}. Below this
temperature the electrons decouple slightly from the base
temperature of our dilution refrigerator; we then use the
resistivity of the Hall bar as the thermometer, assuming that the
$\ln(T)$ law holds down to the lowest temperature.

\begin{figure}
\begin{center}
\includegraphics[width=6.5cm]{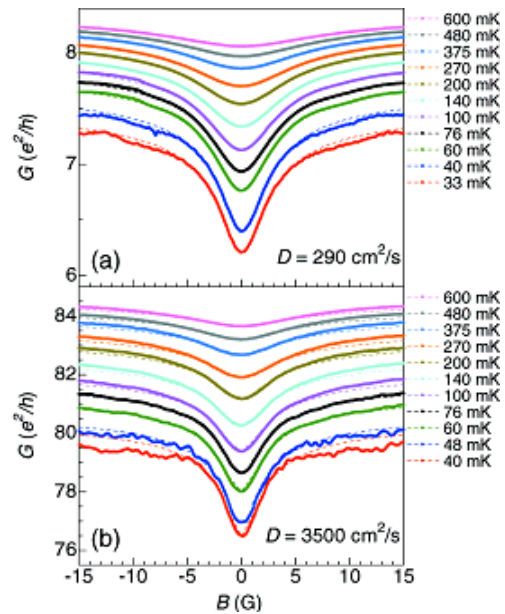}
\caption{(Color online) Magnetoconductance curves of 2DEG wires with
$w = 1130$ nm for (a) $D = 290$ cm$^{2}/$s ($l_{e} =
0.34$ $\mu$m) and (b) 3500 cm$^{2}/$s (4.0 $\mu$m) at different
temperatures. The broken lines show theoretical fits to weak
localization theory for (a) the diffusive and (b) the semi-ballistic
regimes.} \label{fig1}
\end{center}
\end{figure}

Depending on the number of impurities different scattering regimes
are accessible: the ballistic regime ($l_{e} > W,L$), the semi-ballistic 
regime ($W< l_{e} < L$) and the diffusive regime ($l_{e} <
W,L$). The latter two regimes have been explored in this work.

{\em Diffusive regime}: When $l_e < W$, 
the scattering of the electrons is diffusive. 
In this regime, the weak localization
quantum correction to the conductance is given
by~\cite{AAK_82,gilles_book}
\begin{eqnarray}
\Delta G(B)=-2\frac{e^{2}}{h}\frac{N}{L}
\left\{\frac{1}{\sqrt{\frac{1}{L_{\phi}^{2}}+\frac{w^{2}}{3l_{B}^{4}}}}
\right\},
\label{eq_hln}
\end{eqnarray}
where ${e^{2}}/{h}$ is the quantum of conductance, ($e$ is the
charge of the electron, $h$ the Planck's constant), 
$L_{\phi}$ the phase coherence length, $N$ the number
of wires in parallel ($N = 20$ in the present work) and $w$ is the
effective width of the wires, different from the lithographic width
$W$ due to lateral depletion effects inherent to the etching
process. We therefore determine $w$ by fitting the
magnetoconductance at a given temperature and diffusion coefficient:
for a lithographic width $W=$ 1500 nm we obtain $w =$ 1130 nm. This
value is then kept fixed for the entire fitting procedure.

{\em Semi-ballistic regime} ($W<l_{e}$): In this regime it is
necessary to take into account specular reflections on the boundary
of the wires and flux cancelation effects. The weak localization
correction has been calculated by Beenakker and van Houten
(BvH)~\cite{Beenakker_prb_88} and are given below:
\begin{eqnarray}
\Delta G(B)=-2\frac{e^{2}}{h}\frac{N}{L}
\left\{\frac{1}{\sqrt{\frac{1}{L_{\phi}^{2}}+\frac{1}{D\tau_{B}}}}
-\frac{1}{\sqrt{\frac{1}{L_{\phi}^{2}}+\frac{1}{D\tau_{B}}+\frac{2}{l_{e}^{2}}}}
\right\}.
\label{eq_bvh}
\end{eqnarray}
In the magnetic scattering time $\tau_{B}$, both the
\textquotedblleft low\textquotedblright~field and \textquotedblleft
high\textquotedblright~field regimes are taken into account
\textit{via} the expression 
\begin{eqnarray}
D\tau_{B} = D\tau_{B}^{\rm low}+D\tau_{B}^{\rm high}\nonumber =
\frac{9.5}{2}\frac{l_{B}^{4}l_{e}}{w^{3}} +
\frac{4.8}{2}\frac{l_{B}^{2}l_{e}^{2}}{w^{2}}. \nonumber
\label{eq_tau_B}
\end{eqnarray}
As $l_e$ is known from independent measurement on the Hall bar
sample, $L_{\phi}$ remains the only fitting parameter for both
regimes.

\begin{figure}
\begin{center}
\includegraphics[width=6.5cm]{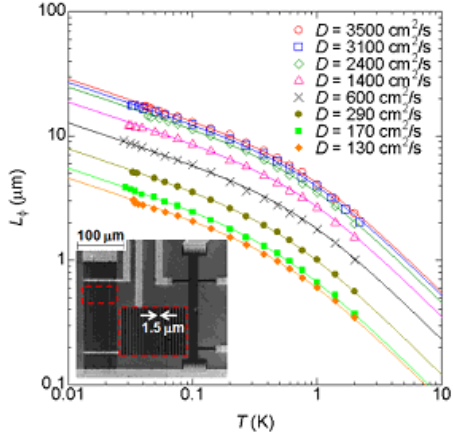}
\caption{(Color online) Phase coherence length $L_{\phi}$ for
different $D$ as a function of $T$ ($w=1130$ nm). The solid
lines are the best fits with Eq.~(\ref{eq_tau_phi}) adding an extra term $\propto T^{-1}$ for high energy processes~\cite{Noguchi_JAP_96}. The open and
closed symbols are used for the semi-ballistic and diffusive
regimes, respectively. The inset shows 2DEG wires with $W=1500$ nm
and a Hall bar.} \label{fig2}
\end{center}
\end{figure}

Typical magnetoconductance curves for the diffusive and
semi-ballistic samples with diffusion coefficients $D=290$ and 3500
cm$^2$/s are displayed in Figs.~\ref{fig1}(a) and~\ref{fig1}(b),
respectively. Fitting the experimental data to Eqs. (\ref{eq_hln})
and (\ref{eq_bvh}), dashed lines in Fig. 1, we can extract 
the phase coherence length $L_{\phi}$ for each temperature. Note, however,
that these expressions are valid only in a regime where the magnetic
length $l_{B}=\sqrt{\hbar/e|B|}$ is larger than the effective width
of the wire $w$; we thus restrict our magnetic field range for
fitting procedure to a field range of $\pm$ 5 G. The observed
deviations above 5 G in Fig.~\ref{fig1}(b) are consistent with the
condition of validity of the BvH expression~\cite{reulet_95}.

The temperature dependence of the phase coherence length $L_{\phi}$
is displayed in Fig.~\ref{fig2} for samples with several different diffusion
coefficients. At low temperatures the dominant decoherence mechanism
is due to electron-electron interactions and the phase coherence
time for quasi-1D diffusive wires is given
by~\cite{AAK_82,gilles_book}
\begin{eqnarray}
\frac{1}{\tau_{\phi}}=\frac{D}{L_{\phi}^{2}}&=&aT^{2/3} \label{eq_tau_phi} \\
&\equiv& \alpha_{\rm AAK}D^{-1/3}T^{2/3} \label{eq_AAK}
\end{eqnarray}
where $\alpha_{\rm AAK}= 1/2(k_{B}\pi/wm^{*})^{2/3}$ with $k_{B}$
the Boltzmann constant and $m^{*}$ the effective mass of the
electron. As shown in Fig.~\ref{fig2}, for all the investigated
samples, $L_{\phi}(T)$ is described very well by
Eq.~(\ref{eq_tau_phi}), where $a$ is treated as a fitting parameter
($a_{\rm exp}$), and adding an extra term $\propto T^{-1}$ for
\textquotedblleft high\textquotedblright~energy
processes~\cite{Noguchi_JAP_96}.

\begin{figure}
\begin{center}
\includegraphics[width=5.6cm]{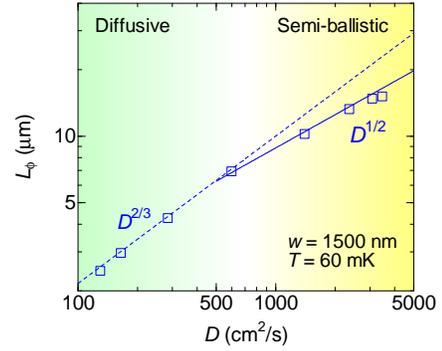}
\caption{(Color online) Phase coherence length $L_{\phi}$ as a
function of $D$ ($w=1130$ nm) at a temperature of $T = 60$ mK.
The solid and broken lines represent $D^{1/2}$ and $D^{2/3}$ laws,
respectively.} \label{fig3}
\end{center}
\end{figure}

\begin{figure}
\begin{center}
\includegraphics[width=6.5cm]{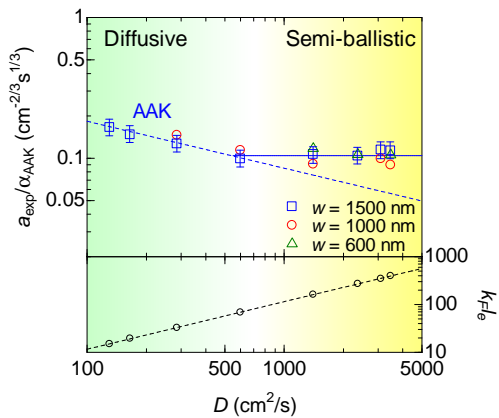}
\caption{(Color online) Top figure shows the experimental
coefficients $a_{\rm exp}$ of Eq.~(\ref{eq_tau_phi}) scaled by
$\alpha_{\rm AAK}$ as a function of $D$. The dashed line represents
$D^{-1/3}$. The experimental values of $k_{F}l_{e}$ are shown in
the bottom part of the figure.} \label{fig4}
\end{center}
\end{figure}

We can go further in the analysis by looking at the $L_{\phi}$ dependence
as a function of $D$. Let us remind that the
diffusion coefficient reflects the strength of the electron-electron
interaction~\cite{AAK_82,gilles_book} and thus controls the low
temperature behavior of the dephasing rate. In Fig.~\ref{fig3}, we
plot $L_{\phi}$ as a function of $D$ at a temperature of 60 mK. One
clearly observes two different regimes: For small $D$, $L_{\phi}$
varies as a power law $D^{\gamma}$ with $\gamma\approx 2/3$, whereas
for large $D$, $\gamma\approx 1/2$. In order to compare more
precisely the $D$ dependence of the prefactor $a$ in
Eq.~(\ref{eq_tau_phi}) with theoretical predictions, we have plotted
in Fig.~\ref{fig4} the experimental parameter $a_{\rm exp}$
extracted from the fits depicted on Fig.~\ref{fig2}, normalized by
the theoretical prefactor $\alpha_{\rm AAK}$ and as a function of
$D$. To compute $a_{\rm exp}$ we use the relation $L_{\phi} =
\sqrt{D\tau_{\phi}}$, which is valid in the diffusive limit as well
as in the semi-ballistic one when reflections on the boundaries are
specular (as is the case for our
samples)~\cite{Beenakker_prb_88,mailly_prb_97}. Furthermore, we have
checked that the resistivity for all our samples is linear with $D$:
this fact also justifies the above assumption. In the diffusive
regime (small $D$), we obtain that the parameter $a_{\rm
exp}/\alpha_{\rm AAK}$ follows a power law as a function of $D$,
with $a_{\rm exp}/\alpha_{\rm AAK}\propto D{^{-1/3}}$, consistent
with Eq.~(\ref{eq_AAK}). Moreover, the prefactor $a_{\rm exp}$
obtained in this work agrees with Eq.~(\ref{eq_AAK}) in
\emph{absolute} value within $15\%$.

Surprisingly, a new regime appears when we go further in the
\textquotedblleft metallic\textquotedblright~regime: when
$k_{F}l_{e}$ increases [see lower part of Fig.~\ref{fig4}], the
samples become semi-ballistic. In this regime, the parameter $a_{\rm
exp}/\alpha_{\rm AAK}$ becomes independent of the diffusion
coefficient. What could be the origin of this disorder independent
decoherence? One could argue that for such high mobility samples the
scattering on the boundaries becomes dominant: if the boundary
scattering of the electrons is diffusive, the diffusion coefficient
should eventually saturate at $D=1/2 v_F w$ \cite{Beenakker_prb_88}
and could lead to disorder-independent decoherence. This possibility,
however, is easily ruled out. Our samples have been made by shallow
etching and the reflections on the boundaries are mainly
specular~\cite{mailly_prb_97}. In addition, the resistance of all
our samples varies linearly with $D$. On the other hand, it is
remarkable that the temperature dependence of $L_{\phi}$ follows
extremely well the power law predicted in~\cite{AAK_82} even
in this semi-ballistic regime~\cite{aleiner_prb_02}. 
This shows that the role of the
temperature in the decoherence process is well captured in the
theory; the role of the disorder, on the other hand, has to be
reconsidered to describe properly our experimental observations. 
To our knowledge, there is only one theoretical attempt to 
consider this regime where the amplitude of the weak localization is 
reduced with increasing $l_{e}$~\cite{schmid_JLTP_87}, but
this calculation has been performed (numerically) for a two-dimensional 
system and only at two typical values of $l_{e}$ and hence does not 
allow for a quantitative analysis of our experimental data.  

Finally, let us mention that in the GZ theory, the low temperature
behavior of the decoherence time $\tau_{\phi}$ should strongly
depend on $D$ ~\cite{GZ_prl_98}. In particular $\tau_{\phi}(T)$ is
expected to saturate at zero temperature at a finite value
$\tau^{0}_{\phi}$ which itself should depend very strongly on $D$.
Such a saturation is not observed for any of our samples for all
diffusion coefficients investigated. This is corroborated by the
fact that for decoherence measurements in our Hall bars with even
lower diffusion coefficients (71 and 46 cm$^{2}/$s) no saturation is
observed. On the contrary, for similar values of diffusion
coefficient in metals, low-temperature saturation of $\tau_{\phi}$
is frequently observed~\cite{mohanty_prl_97,mallet_prl_06}: this
strongly suggests that the low temperature saturation of \tauphi is
\emph{not} intrinsic and hence inconsistent with the GZ theory.

In conclusion, we have measured the disorder dependence of
the phase coherence time in mesoscopic wires made from a two
dimensional electron gas by varying the electronic diffusion
coefficient over 2 orders of magnitude using an original ion
implantation technique. We show that in the diffusive regime, the
phase coherence time follows a power law as a function of the
diffusion coefficient, $D^{\alpha}$,
with $\alpha$ close to $1/3$,
consistent with the 
standard model of decoherence proposed in
Ref.~\cite{AAK_82}. When increasing the diffusion coefficient, the
parameter $k_{F}l_{e}$ increases and samples become semi-ballistic;
we then observe a new regime in which $\tau_\phi$ is
independent of the diffusion coefficient.

\acknowledgments We acknowledge helpful discussions with G.
Montambaux, S. Kettemann, A. D. Zaikin, S. Florens, C. Strunk, and R.
Whitney. Y.~N. acknowledges financial support from the
\textquotedblleft JSPS Research program for Young
Scientists\textquotedblright. This work has been supported by the
European Commission FP6 NMP-3 project 505457-1 \textquotedblleft
Ultra 1D\textquotedblright~and the \textsl{Agence Nationale de la
Recherche} under the grant ANR PNano \textquotedblleft
QuSpin\textquotedblright.

\end{document}